# Dissipative Particle Dynamics for Directed Self-Assembly of Block Copolymers

Hejin Huang[1, a)] and Alfredo Alexander-Katz[1, b)]

## AFFILIATIONS

[1]Department of Materials Science and Engineering, Massachusetts Institute of Technology, Massachusetts 02139, USA

a)Electronic mail: hejin@mit.edu
b)Electronic mail: aalexand@mit.edu

## ABSTRACT

The dissipative particle dynamics (DPD) simulation method has been shown to be a promising tool to study self-assembly of soft matter systems. In particular, it has been used to study block copolymer (BCP) self-assembly. However, previous parametrizations of this model are not able to capture most of the rich phase behaviors of block copolymers in thin films nor in directed self-assembly (chemoepitaxy or graphoepitaxy). Here we extend the applicability of the DPD method for BCPs to make it applicable to thin films and directed self-assembly. Our new reparametrization is able to reproduce the bulk phase behavior, but also manages to predict thin film structures obtained experimentally from chemoepitaxy or graphoepitaxy. A number of different complex structures, such as bilayer nanomeshes, 90° bend structures, circular cylinders/lamellae and Frank-Kasper phases directed by trenches, post arrays or chemically patterned substrate have all been reproduced in this work. This reparametrized DPD model should serves as a powerful tool to predict BCP self-assembly, especially in some complex systems where it is difficult to implement SCFT.

## I. INTRODUCTION

With the advancement of modern technology, intricate designs of nanostructures with nanoscale feature sizes are needed for various applications. Self-assembly of block copolymers (BCPs) offers a feasible route to fabricate such nanopatterns with feature sizes ranging from 5-500nm[1–11]. Block copolymers are comprised of two or more chemically distinct subchains, which are linked together by covalent bonds. Upon annealing, these different blocks phase separate into various microstructures, such as spheres, cylinders or lamellae[12]. Over the past two decades, researchers have developed different techniques to direct the self-assembly process to fabricate new, complex and device-oriented nanopatterns[13–20].

In order to predict the phase structures and explore the impact of various process parameters, a number of methods have been developed to simulate BCP self-assembly. Among these different methods, field-based simulations such as self-consistent field theory (SCFT) have been successful in describing the final structure at equilibrium[21–25]. However, SCFT only gives good prediction for different phenomenon near mean field conditions and in equilibrium, while the analysis and prediction far from equilibrium are not reliable. Extensions using a complex langevin formalism have allowed to include fluctuations in this approach[26–28]. Furthermore, for some complex architectures, such as cross-linked polymers or nanoparticles, field theoretic methods pose serious challenges in the implementation and run time.

Other advances have incorporated a particle-based approach to solving the propagator component for the SCFT method, but sample the Hamiltonian still in the density degrees of freedom. This chain in field approach has been used to model many different systems of BCPs[29–33]. However, given the nature of the chains it is not possible to do non-equilibrium simulations with this method either, albeit there have been some advances in this respect by using slip-links to constraint the polymers and mimic a real environment[34].

Compared to field-based simulations, particle-based simulations give a much more intuitive feel of different experiment conditions at different stages of self-assembly process. In the BCP self-assembly field, however, standard particle-based simulations have only shown limited capability in reproducing experimental results of BCP. These simulations manage to reproduce bulk phase diagrams but they fail in most thin film structures, especially the complex structures obtained by chemoepitaxy or graphoepitaxy. Because of the long self-assembly time, researchers used small chains and compressible systems in particle-based simulation to accelerate the dynamics; however, this introduces artifacts that strongly affect the results. This problem is quite serious in previous DPD studies, where each linear BCP is represented by only 10 beads. In fact, DPD has the capability of performing very fast simulations, but current parametrizations have hindered its advance. In this work, we implement a reparametrized DPD simulation model with larger chains, different connectivity, and higher density. This revised DPD model is able to reproduce some of the most complicated uniform phases observed in diblock thin films, and we expect it to be an effective tool to predict experimental results of both bulk and thin film structures in and out of equilibrium.

## II. METHODS

### A. Dissipative Particle Dynamics (DPD)

The setup for DPD in this work is modified from the previously reported DPD method[34]. In particular, each linear diblock copolymer is represented by 20 beads (compared to 10 beads for previous studies)[34–36]. Each bead can be of A and B types, as shown in Figure 1(a). In this paper, $A_xB_y$ refers to the linear BCP with each chain consisting of x beads of type A and y beads of type B in the DPD model. For example, $A_4B_{16}$ corresponds to a linear diblock copolymer with the A composition being 20% of the total volume.

As is a custom in DPD, we use reduced units throughout this paper: thermal energy kT is set to 1 and the mass of DPD particles of polymers is also set to 1. There other two parameters that are varied in the simulation are: volume ratio of A and the graphoepitaxy/ chemoepitaxy configurations. The beads within the same polymer are interconnected with harmonic bonds in a linear fashion. The force field of the harmonic bond is given by:

$$F_{i,i+1}^{bond} = K(r_{i,i+1} - r_0)\hat{r}_{ij} \qquad (1)$$

In equation (1), K represents the spring constant while $r_0$ is the equilibrium distance. In previous DPD studies, the parameters of harmonic bonds were $r_0=0$ and K=4. In our case, however, the spring constant is changed to K=50 with the equilibrium distance being 1.0 in dimensionless units. This small change appears to be significant for obtaining the correct thin film morphologies. We believe this is due to the fact that in the previous setup, the chains will tend to "collapse" once they are in the phase separated regime to minimize the bond energy. Fixing the bond length to a nonzero value prevents the chains from collapsing. In any case, since we are working with 20 beads, it is simple to show that in this regime the Fokker-Planck eq. is that of a Gaussian chain within the freely jointed chain model used here.

Besides the harmonic bond, a soft repulsive force exist between two random beads that are close to each other. This force is defined by the equation:

$$F_{ij}^C = \begin{cases} -a_{ij}(1-|r_{ij}|)\hat{r}_{ij} & if\ |r_{ij}| < 1 \\ 0 & if\ |r_{ij}| \geq 1 \end{cases} \qquad (2)$$

In Equation (2), $a_{ij}$ represents the maximum repulsion between particle $i$ and particle $j$, and $r_{ij}$ and $\hat{r}_{ij}$ are the distance and unitary vector between the two particles respectively. Based on previously reported literature, the repulsion parameter for the same particle type is[37]:

$$a_{ii}\rho = 75kT \qquad (3)$$

where ρ is the bead density, $k$ is the Boltzmann constant and $T$ is the temperature. In our simulation, we choose the density to be $\rho = 5$, which gives $a_{ii} = 15kT$. The Flory-Huggins parameters χ between two different species is mapped to the repulsion parameter by the equation[37]:

$$a_{ij} \approx a_{ii} + 1.45\chi_{ij} \qquad (4)$$

Besides the soft repulsive potential, the two other forces are thermal fluctuations and drag forces. These two forces are defined by the following equations[37]:

$$F_{ij}^R = \sigma w^R(r_{ij})\theta_{ij}\hat{r}_{ij}\zeta/(\delta t)^{1/2} \qquad (5)$$

$$F_{ij}^D = \frac{1}{2}\sigma^2 \left(w^R(r_{ij})\right)^2 /kT(v_{ij}\cdot\hat{r}_{ij})\hat{r}_{ij} \qquad (6)$$

The ζ in equation (5) is a random variable with zero mean and variance one, and $w(r) = (1-r)$ for $r < 1$ and $w = 0$ for $r > 1$. In this paper, the time step is set to 0.015 while the noise factor σ = 0.10. All simulations are run for at least 1000000 time steps, certain simulations have been run for 2000000 steps. Note that the value of the noise factor used here is significantly different from the value reported in previous literature (4.65). It is found that with this reparametrized DPD method is able to reproduce nearly all experimental results obtained from BCP self-assembly. The noise previously used in simulation (σ >4.00) is too big to form nonregular-shaped nanostructures seen in experiments. The interaction potential (unless otherwise noted) is set to $a_{AB} = 25.9$, corresponding to $\chi N^{eff} = 60$ (for N=20). For the bulk simulation in Section III A, the box size is $L_x = L_y = L_z = 30$. For thin film simulation, unless otherwise stated, the length and width of the box is $L_x = L_y = 60$ while the height of the box is $L_z = 10$ for single-layer structures.

## B. Setup of thin films with chemoepitaxy and graphoepitaxy

*(a) BCP Thin Film*

Self-assembly of BCP thin films is simulated by introducing two additional categories of beads - $w_1$ and $w_2$ with fixed position, as shown in Figure 1(b). The $w_1$ beads on top simulates the air-BCP interface, while the $w_2$ beads at the bottom simulates BCP-substrate interface. These two beads have higher density – approximately $2.5\rho$ in order to prevent penetration of BCP into the interfaces. The interaction parameters are based on observed experimental conditions, e.g. PMDS has a lower surface energy than PS. For BCPs that form in-plane structures (such as PS-PDMS), the air interface is attractive to minority block A while the substrate is attractive to the majority block B. Therefore, the interaction parameters $a_{ij}$ is set as followed:

$$a_{Aw_1} = a_{Bw_2} = a_{ii} \tag{7}$$

$$a_{Aw_2} = a_{Bw_1} = a_{AB} \tag{8}$$

For BCPs that form standing up structures (such as PS-PMMA), both the upper and lower interfaces are neutral to both subchains. In the simulation, the interaction parameter between $w_1$ and $w_2$ with other particles are:

$$a_{Aw_1} = a_{Bw_1} = a_{Aw_2} = a_{Bw_2} = a_{ii} \tag{9}$$

*(b) Chemoepitaxy*

For chemoepitaxy, the substrate surface is chemically patterned by two categories of brush layers at different regions: one is attractive to subchain A while the other one is attractive to subchain B. By varying the pattern of the brush layer on the substrate surface, one can control the final self-assembled structure.

In the DPD, chemoepitaxy can be simulated by introducing three categories of beads with fixed positions: two in the lower boundary ($w_1$ and $w_2$) and one in the upper boundary ($w_3$) as shown in Figure 1(c). The interaction parameters are set as follows:

$$a_{Aw_1} = a_{Bw_2} = a_{Aw_3} = a_{Bw_3} = a_{ii} \tag{10}$$

$$a_{Aw_2} = a_{Bw_1} = a_{AB} \tag{11}$$

By varying locations of $w_1$ and $w_2$ beads, one can change the chemical patterns of the substrate.

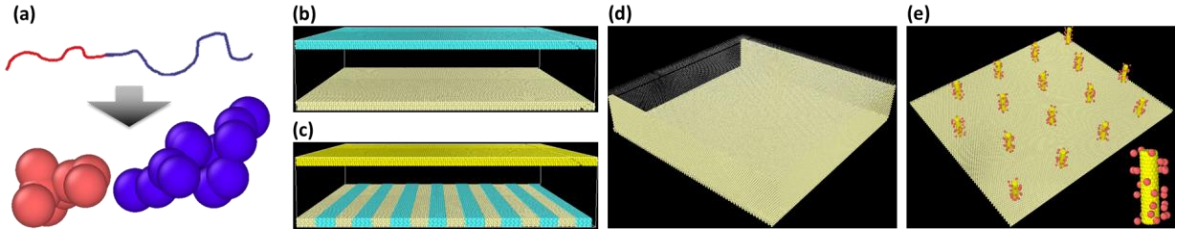

Figure 1 Simulation setup of for DPD of linear diblock copolymers. In (a) we sketch the polymer. For thin film simulations, two additional substrates are necessary as shown in (b). The substrates can have a pattern and this corresponds to chemoepitaxy as shown in (c). Finally, we also consider graphoepitaxy in the form of trenches (d) and post arrays with brush layers (e).

*(c) Graphoepitaxy*

Graphoepitaxy serves as one of the most commonly used methods to fabricate uniform and complex nanostructures from self-assembly of BCP. Different graphoepitaxial templates, such as post arrays, substrate modulation, or confinement by trenches have enabled researchers to fabricate different 2D nanostructures. In this simulation work, two commonly used graphoepitaxial templates are demonstrated: trenches and post arrays.

Trenches are simulated by introducing beads forming sidewalls with fixed positions, as shown in Figure 1(d). The bead type of these sidewalls is the same as the bottom substrate. These sidewalls serve as heterogeneous nucleation site for self-assembly of BCP, which in theory will lead to formation of uniform structures.

The post arrays are simulated by introducing an additional type of beads forming posts on the substrates shown in Figure 1(e). The beads of the post arrays are neutral to both subchains of BCP. However, by introducing a brush layer to these post arrays, one can directed the self-assembly process to form a uniform final structure. In simulations, either A or B homopolymers are attached to the post arrays by harmonic bonds, as shown in Figure 1(e). By controlling the post symmetry and inter-post distance, one can control the final assembled structures.

## III. RESULTS & DISCUSSION

### A. Bulk Structure

After setting up the 20-bead system, the phase diagram for bulk BCP self-assembly is constructed by the DPD simulation method through varying the volume ratios (step size: 5%) and χN (step size: 10). The phase diagram by DPD simulation of 20-bead system is shown in Figure 2(a). This symmetric phase diagram matches very well both the mean-field model and the experimental result.

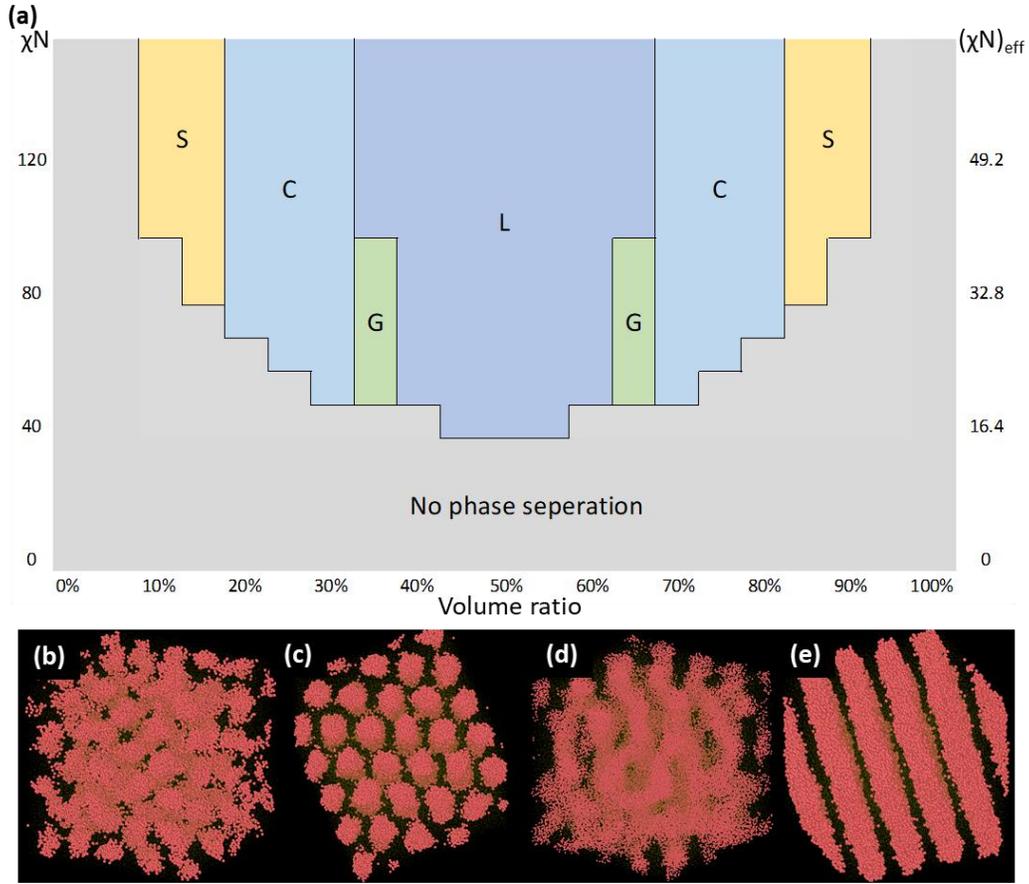

Figure 2 (a) Phase diagram of linear diblock copolymer from DPD simulation; (b)-(e) snapshots of different phase structures (b) spheres, (c) cylinders, (d) gyroid, and (e) lamellae. All the bulk phase structures have been reproduced

Comparing to the mean-field phase diagram obtained from self-consistent field theory (SCFT), the DPD results have the same phase behavior: four different structures - Spheres, cylinders, gyroid and lamellae are formed at different volume ratios. Snapshots of these phase structures are shown in Figure 2(b)-(e). It is noted that at the volume ratio of 35%, a gyroid structure is observed at small χN while lamellae is formed when χN is larger than a critical value, which matches with the phase behavior obtained from SCFT and experiments. The only deviation is the critical χN where order-disorder transition occurs when the volume ratio is 50%: the critical χN of DPD simulation is 40, while the critical χN of SCFT and experiments are 10.5 and 16 respectively[12,38].

The main possible cause for deviations is due to fluctuations with a finite size chain with N=20. This has been discussed by Groot et al. in his pioneering work of using DPD to simulate diblock copolymers[34]. In short, by simulating each diblock copolymer with only 20 beads, it artificially increases the importance of fluctuation of each beads, which helps to stabilized isolated micelles instead of certain continuous structures. The effective χ parameter of a system with shorter polymer has been proposed to be:

$$(\chi N)_{eff} = \frac{10.5}{10.5 + 41.0 N^{-1/3}} \chi N = \frac{\chi N}{1 + 3.9 N^{-1/3}} \qquad (12)$$

The critical $(\chi N)_{eff}$ calculated from Equation 12 is 16.4, which matches well the experimental value and is above the SCFT value (see Figure 2). In conclusion, our reparametrized DPD model manages to reproduce phase behavior of linear diblock copolymers quantitatively. In the following sections, we will mainly focus on DPD simulation of thin film structures and directed self-assembly.

**B. Thin Films**

Thin film self-assembly constitutes a very simple form of directed self-assembly. For thin films, the BCP self-assembles within the confinement of upper and lower interfaces that serve as guiding boundaries for the assembly process. The interaction parameters between the BCP and the two interfaces are set as in Equation (7) and (8) (similar to a PS-PDMS experimental system). After self-assembly, a thin film of A is formed at the BCP-air interface while the pattern of A lies underneath the thin film. Figure 3(a)-(h) are top views of the self-assembled patterns at the mid plane when the volume ratio (f) is between 10% and 40%.

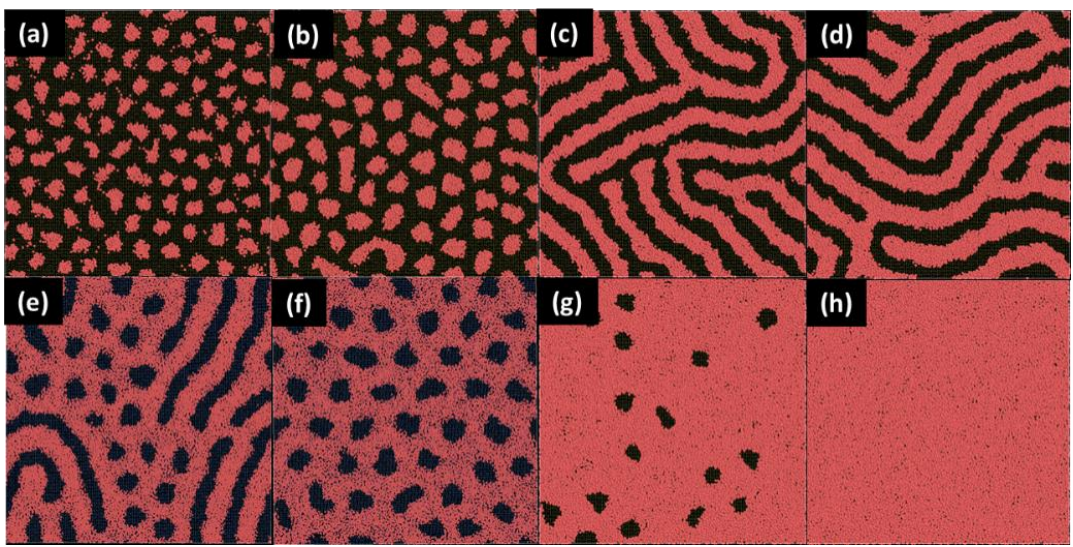

Figure 3 Structures obtained from self-assembly of thin film BCP of (a) $A_2B_{18}$, (b) $A_3B_{17}$, (c) $A_4B_{16}$, (d) $A_5B_{15}$, (e) $A_6B_{14}$, (f) $A_7B_{14}$, (g) $A_7B_{13}$, (h) $A_8B_{12}$. These structures obtained from DPD matches with experimental result.

As shown in Figure 3, different thin film structures, such as hexagonally packed spheres, fingerprint, perforated lamellae and lamellae, have been obtained from DPD simulation. The volume ratio of each structure is very close to the studies from SCFT and experiments[39,40]: spheres – 10% & 15%, cylinder – 20% & 25%, lamellae – 40%. When the volume ratio is 30% and 35%, intermediate structures between two phases are obtained (30% - combination of cylinders and perforated lamellae; 35% - combination of perforated lamellae and lamellae). These transition structures are also physical and have been commonly observed in experiments when the volume ratio is close to the transition value between the two phases[41]. Perforated lamellae with six-fold symmetry is the only phase that is not obtained in the 20 beads system because its equilibrium volume ratio lies somewhere between 30% and 35%. Such structure could be achieved by extending to 21 beads system with a volume ratio of 33.3% ($A_7B_{14}$), as shown in Figure 3(f).

It is noted from Figure 3 that the results from the DPD simulation appear to be much noisier compared to SEM images obtained from experiments: instead of a perfect spherical shape, the spheres in DPD are stretched in Figure 3(a) & (b) due to fluctuations; the boundaries of the cylinders in Figure 3(c) & (d) also appear wavy. Such noise could be reduced by using longer chains, which uses more beads to represent a BCP (i.e. making the result closer to mean field conditions). However, introducing more beads also introduces a higher computational cost. In this paper, thus, despite the noise it introduces, the 20-bead system manages to predict directed self-assembly of block copolymers accurately. The simulation results in the following sections are all based on the 20-bead copolymer.

**C. Graphoepitaxy**

*(a) Trenches*

Trenches are one of the most commonly used techniques to direct BCP self-assembly. Because the feature size of the BCP self-assembled structure (sub 10nm) can be much smaller than the resolution limit of photolithography, researchers usually create relatively large trenches by photolithography, and then use these trenches to direct the BCP self-assembly to obtain uniform nanostructures with sub 10nm feature size[16,42–48]. Different nanostructures such as parallel cylinders/lamellae with long-range order[44,45,47,49], and circular cylinders[15] have been achieved by controlling the trench shape.

Figure 4(a)-(d) shows the simulation results for a cylinder-forming BCP ($A_5B_{15}$) using a non-neutral interface and different trench shapes. Without the trenches, the fingerprint structure with no long-range order is formed, as shown in Figure 4(a). After introducing the parallel sidewalls, the cylinders start to align to the sidewalls, which gives parallel structures with long-range order, as shown in Figure 4(b). According to previous research, the sidewalls act as heterogeneous nucleation sites, which leads to formation of uniform structures. Figure 4(c) & (d) show the self-assembled structures with circular and square trenches: complex nanostructures, such circular cylinders and ladder-shaped patterns with T-junctions are obtained. The structures predicted by DPD simulation match well the experimental results in literature[50,51].

Figure 4(e)-(h) show the simulation result of the system with different trench shapes, lamellae-forming BCP ($A_{10}B_{10}$) and non-neutral interface. Compared with Figure 4(a)-(d), the standing-up lamellae gives similar structures with parallel and circular trenches, but a different structure with square trench. Instead of T-junction, parallel lamellae with 90° bend are formed, as shown in Figure 4(h). This is because formation of T-junction has higher energy penalty for lamellae compared to cylinders. The same phenomenon has been observed in the experiments[52,53].

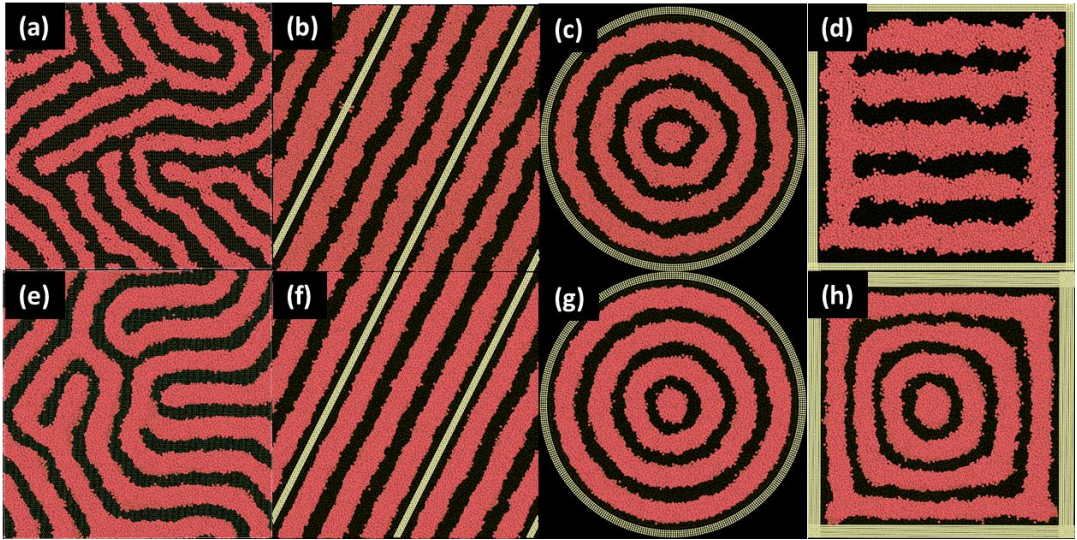

Figure 4 Diblock self-assembly in thin films. (a)-(d) $A_5B_{15}$ cylinder self-assembled structure within different substrates under non-neutral top and bottom interfaces ($a_{Aw_1} = a_{Bw_2} = a_{ii}$ and $a_{Aw_2} = a_{Bw_1} = a_{AB}$): (a) simple thin film, (b) parallel sidewalls/trench, (c) circular trench, (d) square trench; (e)-(h) $A_{10}B_{10}$ lamellae self-assembled structure within different substrates neutral top and bottom interfaces ($a_{Aw_1} = a_{Bw_1} = a_{Aw_2} = a_{Bw_2} = a_{ii}$): (e) simple thin film, (f) parallel sidewalls, (g) circular trench, (h) square trench.

*(b) Post Arrays*

Introducing post arrays is another method to direct the BCP self-assembly process for a desired pattern. These post arrays are fabricated by electron beam lithography, and their feature size can easily go down to sub 5nm. Brush layers, which attracts certain block of the BCP while repelling the other, is introduced to the post arrays to guide the self-assembly process[54]. Owing to its small feature size, post arrays have enabled researchers to have exquisite control of the final self-assembled structures. Over the past decade, researchers have managed to fabricate different nanostructures such as well-aligned cylinders with controlled orientation[14], uniform hexagonally packed spheres[54], bilayer nanomeshes[19], perforated lamellae with non-local symmetry[55] and other complex asymmetric structures[14,24,56]. A number of different nanostructures have been demonstrated in this section.

Figure 5(a)-(d) are the simulation results of the system with square post arrays, cylinder-forming BCP ($A_5B_{15}$) and non-neutral interface. Well-aligned cylinders with different orientations could be obtained by varying the inter-post distances. When the inter-post distance matches the natural periodicity of the cylinders, the cylinders follows the direction of the side of the square post arrays, as shown in Figure 5(e) & (f). When the inter-post distance is approximately $\sqrt{2}$ times as large as the natural periodicity, the cylinders align to the diagonal of the square post arrays, as shown in Figure 5(g) & (h). The differences between Figure 5(e) & (f) and between Figure 5(g) & (h) are whether the brush layer is majority or minority block of BCP. This determines whether the self-assembled cylinders passes through the post arrays or not. These prediction matches with experimental result reported in literature[14].

Figure 5(e)-(h) are the simulation results of the system with square post arrays, cylinder-forming BCP ($A_{15}B_5$) and neutral interface. The brush layer of these posts is the minority block (block B).
9

The neutral interface leads to formation of cylinders with out-of-plane orientation. Different non-local symmetry could be obtained by varying the inter-post distance. When the inter-post distance is small, cylinders forms only at the locations of the post arrays, which has square symmetry. As the inter-post distance increases, cylinders start to show up in other locations. Figure 5(f) & (h) shows the two different inter-post distances where cylinders retain the square symmetry, with either one or four cylinders in the center of every four posts. When the inter-post distance is between the value taken in Figure 5(f) & (h), a Frank-Kasper phase is formed, where 2D diamond structures appear, as highlighted by the dash lines in Figure 5(g). These non-trivial different symmetries predicted by DPD simulation match exactly with the experimental result from our group[55].

Figure 5(i) & (j) are the simulation results of the cylinder-forming BCP ($A_5B_{15}$) in a rectangular post array with non-neutral upper and bottom interfaces. The brush layer of these posts is the majority block (block B). The DPD simulation clearly reproduces the some experimental results from previously reported double layers in graphoepitaxial templates[19].

Figure 5(k) & (l) are the simulation results of the system with hexagonal post arrays with neutral interface. Uniform hexagonally packed standing cylinders with long-range order are formed, as expected. Structures of different orientations can be obtained with different inter-post distances: cylinders arranged in the (1 1) phase with respect to the post arrays are formed at inter-post distance (L=13.0) shown in Figure 5(k), while cylinders in the (2 0) phase are obtained at L= 15.5 (Figure 5(l)) is agreement with experiments doing density multiplication[54].

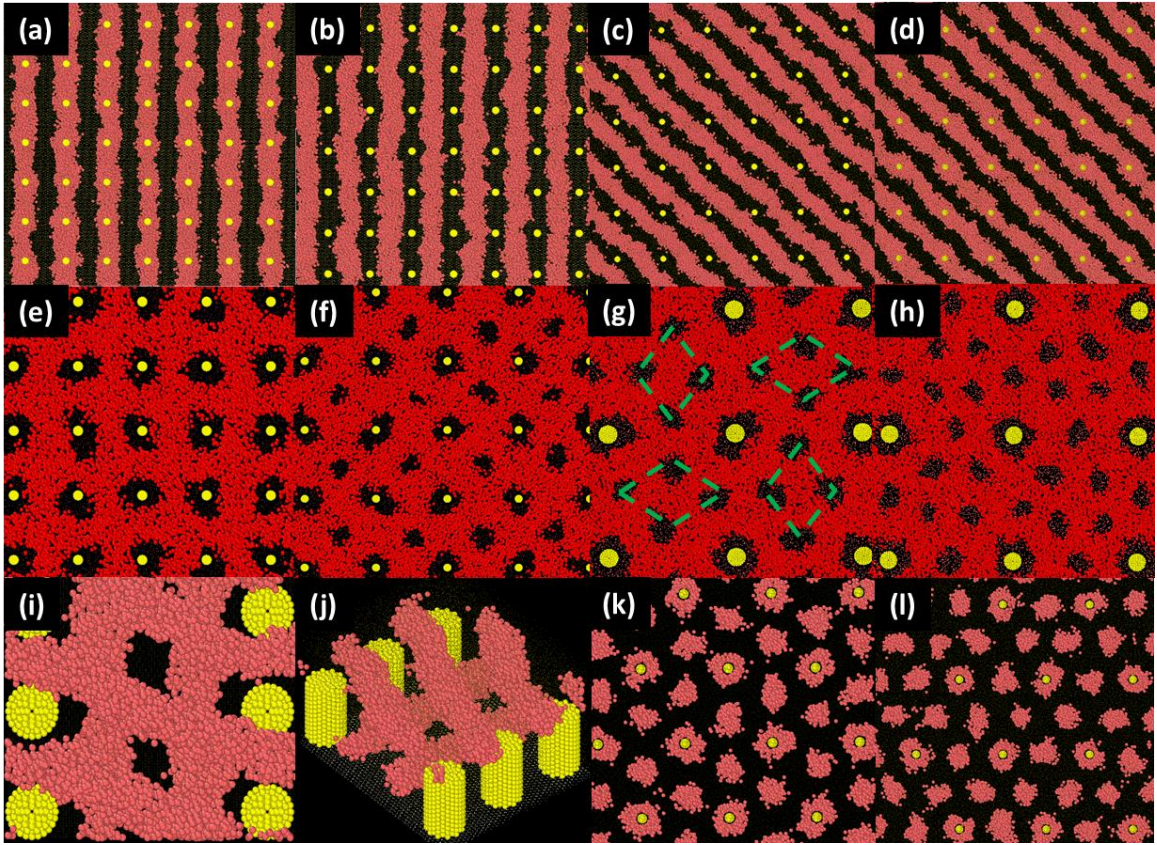

Figure 5 Phases of cylinder forming $A_4B_{16}$ BCPs in different square post arrays. (a)-(d) within non-neutral interfaces ($a_{Aw_1} = a_{Bw_2} = a_{ii}$ and $a_{Aw_2} = a_{Bw_1} = a_{AB}$), (e)-(h) within neutral interfaces ($a_{Aw_1} = a_{Bw_1} = a_{Aw_2} = a_{Bw_2} = a_{ii}$); (i)-(j) thick thin films within rectangular post arrays with non-neutral interfaces; (k)-(l) density multiplication of cylinder BCPs in hexagonal post arrays with neutral interfaces. Note that all these results are in excellent agreement with experimentally observed results[14,19,54,55]

## D. Chemoepitaxy

Other than graphoepitaxy, another commonly used method to achieve complex nanostructures is chemoepitaxy. By introducing chemically nanopatterned surfaces, which direct the BCP self-assembly process, uniform nanostructures such as parallel lamellae with long-range order[57,58], and nonregular-shaped nanostructures such as bends[17,52] have been achieved.

Figure 6(a)-(d) show the simulation results for a lamellae-forming BCP ($A_{10}B_{10}$) in different chemical patterns with a neutral top interface. As already seen in Figure 3(e), with neutral interfaces, standing-up lamellae with no long-range order is formed. When the underlying $w_1$ and $w_2$ beads create a parallel pattern that matches the periodicity of the BCP, parallel lamellae with long-range order is formed, shown in Figure 6(a) & (b). Moreover, nonregular-shaped nanostructures containing bends, such as squares and cross, could be achieved by introducing complex chemical pattern on the bottom substrate, as shown in Figure 6(c) & (d). All these structures match with the SEM images of the experimental results previously reported[17,58].

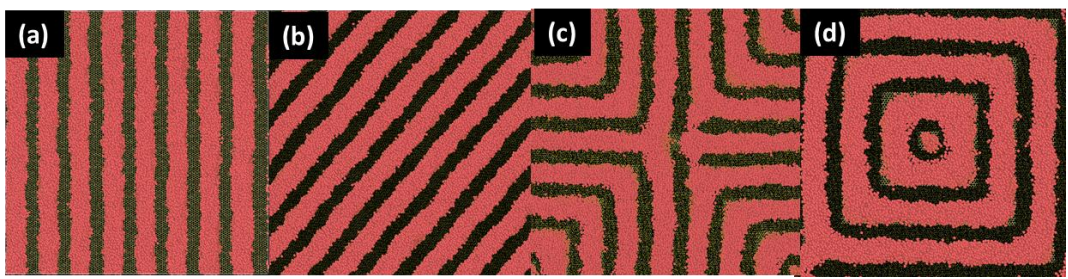

Figure 6 Chemoepitaxy directed patterns formed from a lamellae forming BCP $A_{10}B_{10}$ film: (a) vertically aligned lamellae, (b) 45° lamellae, (c) cross (d) squares with 90° bend structures.

## IV. CONCLUSION

In summary, we have presented a reparametrized DPD simulation model that has the ability to reproduce almost every thin film morphology produced by experiments. This reparametrization relies on small changes to current DPD implementations. In particular, we represent each linear diblock copolymer with 20 beads using freely jointed chain model, increase the density, and lower the temperature. We believe this is a new transferable parametrization that can reproduce both the BCP bulk phase diagram and thin film phase behavior. Moreover, all the complex thin film structures obtained from experiments with trenches, post arrays or chemically patterned substrate have been accurately predicted by this DPD parametrization as well. Previous parametrizations failed dramatically under these later conditions. We believe, thus, that our proposed DPD parmetrization provides an additional powerful tool for experimental design and structure prediction. Since it is a particle-based simulation, the biggest advantages is its easy implementation for a complex system compared to SCFT: systems containing crosslinks and nanoparticles could be easily represented in this particle-based model.

## ACKNOWLEDGMENTS

This work was supported by the U.S. Department of Energy, Office of Basic Energy Sciences, Division of Materials Science and Engineering under Award No. #ER46919.

## REFERENCES

[1] J.G. Kennemur, L. Yao, F.S. Bates, and M.A. Hillmyer, Macromolecules **47**, 1411 (2014).

[2] K. Aissou, M. Mumtaz, G. Fleury, G. Portale, C. Navarro, E. Cloutet, C. Brochon, C.A. Ross, and G. Hadziioannou, Adv. Mater. **27**, 261 (2015).

[3] H.S. Suh, D.H. Kim, P. Moni, S. Xiong, L.E. Ocola, N.J. Zaluzec, K.K. Gleason, and P.F. Nealey, Nat Nanotechnol **12**, 575 (2017).

[4] H.-Y. Tsai, H. Miyazoe, S. Engelmann, B. To, E. Sikorski, J. Bucchignano, D. Klaus, C.-C. Liu, J. Cheng, D. Sanders, N. Fuller, and M. Guillorn, Journal of Vacuum Science & Technology B,

Nanotechnology and Microelectronics: Materials, Processing, Measurement, and Phenomena **30**, 06F205 (2012).

[5] M. Delalande, G. Cunge, T. Chevolleau, P. Bézard, S. Archambault, O. Joubert, X. Chevalier, and R. Tiron, Journal of Vacuum Science & Technology B, Nanotechnology and Microelectronics: Materials, Processing, Measurement, and Phenomena **32**, 051806 (2014).

[6] A.P. Lane, X. Yang, M.J. Maher, G. Blachut, Y. Asano, Y. Someya, A. Mallavarapu, S.M. Sirard, C.J. Ellison, and C.G. Willson, ACS Nano **11**, 7656 (2017).

[7] S. Park, D.H. Lee, J. Xu, B. Kim, S.W. Hong, U. Jeong, T. Xu, and T.P. Russell, Science **323**, 1030 (2009).

[8] C. Choi, J. Park, K.L.V. Joseph, J. Lee, S. Ahn, J. Kwak, K.S. Lee, and J.K. Kim, Nature Communications **8**, 1765 (2017).

[9] A. Noro, Y. Tomita, Y. Shinohara, Y. Sageshima, J.J. Walish, Y. Matsushita, and E.L. Thomas, Macromolecules **47**, 4103 (2014).

[10] C.-G. Chae, Y.-G. Yu, H.-B. Seo, M.-J. Kim, R.H. Grubbs, and J.-S. Lee, Macromolecules **51**, 3458 (2018).

[11] A.L. Liberman-Martin, C.K. Chu, and R.H. Grubbs, Macromolecular Rapid Communications **38**, 1700058 (2017).

[12] F.S. Bates and G.H. Fredrickson, Physics Today **52**, 32 (2008).

[13] S. Ji, U. Nagpal, W. Liao, C.-C. Liu, J.J. de Pablo, and P.F. Nealey, Advanced Materials **23**, 3692 (2011).

[14] J.K.W. Yang, Y.S. Jung, J.-B. Chang, R.A. Mickiewicz, A. Alexander-Katz, C.A. Ross, and K.K. Berggren, Nature Nanotechnology **5**, 256 (2010).

[15] A.T.K. G, S.M. Nicaise, K.R. Gadelrab, A. Alexander-Katz, C.A. Ross, and K.K. Berggren, Nature Communications **7**, 10518 (2016).

[16] C.T. Black and O. Bezencenet, IEEE Transactions on Nanotechnology **3**, 412 (2004).

[17] M.P. Stoykovich, M. Müller, S.O. Kim, H.H. Solak, E.W. Edwards, J.J. de Pablo, and P.F. Nealey, Science **308**, 1442 (2005).

[18] P.W. Majewski, A. Rahman, C.T. Black, and K.G. Yager, Nature Communications **6**, 7448 (2015).

[19] A.T.K. G, K.W. Gotrik, A.F. Hannon, A. Alexander-Katz, C.A. Ross, and K.K. Berggren, Science **336**, 1294 (2012).

[20] M.P. Stoykovich, H. Kang, K.Ch. Daoulas, G. Liu, C.-C. Liu, J.J. de Pablo, M. Müller, and P.F. Nealey, ACS Nano **1**, 168 (2007).

[21] S.W. Sides and G.H. Fredrickson, J. Chem. Phys. **121**, 4974 (2004).

[22] S.W. Sides and G.H. Fredrickson, Polymer **44**, 5859 (2003).

[23] S.-M. Hur, C.J. García-Cervera, E.J. Kramer, and G.H. Fredrickson, Macromolecules **42**, 5861 (2009).

[24] A.F. Hannon, Y. Ding, W. Bai, C.A. Ross, and A. Alexander-Katz, Nano Lett. **14**, 318 (2014).

[25] H. Ceniceros and G. Fredrickson, Multiscale Model. Simul. **2**, 452 (2004).

[26] G.H. Fredrickson, V. Ganesan, and F. Drolet, Macromolecules **35**, 16 (2002).

[27] A. Alexander-Katz and G.H. Fredrickson, Macromolecules **40**, 4075 (2007).

[28] A. Alexander-Katz, A.G. Moreira, S.W. Sides, and G.H. Fredrickson, J. Chem. Phys. **122**, 014904 (2004).

[29] D.Q. Pike, F.A. Detcheverry, M. Müller, and J.J. de Pablo, J. Chem. Phys. **131**, 084903 (2009).

[30] S.-M. Hur, V. Thapar, A. Ramírez-Hernández, G. Khaira, T. Segal-Peretz, P.A. Rincon-Delgadillo, W. Li, M. Müller, P.F. Nealey, and J.J. de Pablo, PNAS **112**, 14144 (2015).

[31] F.A. Detcheverry, G. Liu, P.F. Nealey, and J.J. de Pablo, Macromolecules **43**, 3446 (2010).

[32] F.A. Detcheverry, D.Q. Pike, P.F. Nealey, M. Müller, and J.J. de Pablo, Phys. Rev. Lett. **102**, 197801 (2009).

[33] U. Nagpal, M. Müller, P.F. Nealey, and J.J. de Pablo, ACS Macro Lett. **1**, 418 (2012).

[34] R.D. Groot and T.J. Madden, J. Chem. Phys. **108**, 8713 (1998).

[35] P. Petrus, M. Lísal, and J.K. Brennan, Langmuir **26**, 3695 (2010).

[36] J. Feng, H. Liu, and Y. Hu, Macromolecular Theory and Simulations **15**, 674 (2006).

[37] R.D. Groot and P.B. Warren, J. Chem. Phys. **107**, 4423 (1997).

[38] A.K. Khandpur, S. Foerster, F.S. Bates, I.W. Hamley, A.J. Ryan, W. Bras, K. Almdal, and K. Mortensen, Macromolecules **28**, 8796 (1995).

[39] P. Mansky, P. haikin, and E.L. Thomas, JOURNAL OF MATERIALS SCIENCE **30**, 1987 (1995).

[40] W. Li, M. Liu, F. Qiu, and A.-C. Shi, J. Phys. Chem. B **117**, 5280 (2013).

[41] L. Tsarkova, A. Knoll, G. Krausch, and R. Magerle, Macromolecules **39**, 3608 (2006).

[42] Y.S. Jung and C.A. Ross, Small **5**, 1654 (2009).

[43] W. Lee, R. Ji, C.A. Ross, U. Gösele, and K. Nielsch, Small **2**, 978 (2006).

[44] C.C. Kathrein, W. Bai, J.A. Currivan-Incorvia, G. Liontos, K. Ntetsikas, A. Avgeropoulos, A. Böker, L. Tsarkova, and C.A. Ross, Chem. Mater. **27**, 6890 (2015).

[45] Y.S. Jung and C.A. Ross, Nano Lett. **7**, 2046 (2007).

[46] J.Y. Cheng, A.M. Mayes, and C.A. Ross, Nature Materials **3**, 823 (2004).

[47] D. Sundrani, S.B. Darling, and S.J. Sibener, Nano Letters **4**, 273 (2004).

[48] R.A. Segalman, H. Yokoyama, and E.J. Kramer, Advanced Materials **13**, 1152 (2001).

[49] W. Bai, K. Gadelrab, A. Alexander-Katz, and C.A. Ross, Nano Lett. **15**, 6901 (2015).

[50] H.K. Choi, J.-B. Chang, A.F. Hannon, J.K.W. Yang, K.K. Berggren, A. Alexander-Katz, and C.A. Ross, Nano Futures **1**, 015001 (2017).

[51] H.W. Do, H.K. Choi, K.R. Gadelrab, J.-B. Chang, A. Alexander-Katz, C.A. Ross, and K.K. Berggren, Nano Convergence **5**, 25 (2018).

[52] G.M. Wilmes, D.A. Durkee, N.P. Balsara, and J.A. Liddle, Macromolecules **39**, 2435 (2006).

[53] M.S. Onses, C. Song, L. Williamson, E. Sutanto, P.M. Ferreira, A.G. Alleyne, P.F. Nealey, H. Ahn, and J.A. Rogers, Nature Nanotechnology **8**, 667 (2013).

[54] I. Bita, J.K.W. Yang, Y.S. Jung, C.A. Ross, E.L. Thomas, and K.K. Berggren, Science **321**, 939 (2008).

[55] Y. Ding, K. Gadelrab, K.M. Rodriguez, W. Huang, C.A. Ross, and A. Alexander-Katz, Nature Communications **In press**, (2019).

[56] J.-B. Chang, H.K. Choi, A.F. Hannon, A. Alexander-Katz, C.A. Ross, and K.K. Berggren, Nature Communications **5**, 3305 (2014).

[57] R. Ruiz, H. Kang, F.A. Detcheverry, E. Dobisz, D.S. Kercher, T.R. Albrecht, J.J. de Pablo, and P.F. Nealey, Science **321**, 936 (2008).

[58] S.O. Kim, H.H. Solak, M.P. Stoykovich, N.J. Ferrier, J.J. de Pablo, and P.F. Nealey, Nature **424**, 411 (2003).